\pdfoutput=1
\documentclass[conference]{IEEEtran}
\IEEEoverridecommandlockouts
\usepackage{cite}
\usepackage{amsmath,amssymb,amsfonts}
\usepackage{algorithmic}
\usepackage{graphicx}
\usepackage{textcomp}
\usepackage{xcolor}
\usepackage{booktabs}
\usepackage{float}
\def\BibTeX{{\rm B\kern-.05em{\sc i\kern-.025em b}\kern-.08em
    T\kern-.1667em\lower.7ex\hbox{E}\kern-.125emX}}
\begin{document}

\title{Improving the efficiency of spectral features extraction by structuring the audio files}

\author{\IEEEauthorblockN{Dishant Parikh}
\IEEEauthorblockA{\textit{Computer Engineering} \\
\textit{GCET}\\
Gujarat, India \\
dishant30899@gmail.com}
\and
\IEEEauthorblockN{Saurabh Sachdev}
\IEEEauthorblockA{\textit{Information Technology} \\
\textit{GCET}\\
Gujarat, India \\
saurabh15101998@gmail.com}
}

\maketitle

\begin{abstract}
The extraction of spectral features from a music clip is a computationally expensive task. As in order to extract accurate features, we need to process the clip for its whole length. This preprocessing task creates a large overhead and also makes the extraction process slower. We show how formatting a dataset in a certain way, can help make the process more efficient by eliminating the need for processing the clip for its whole duration, and still extract the features accurately. In addition, we discuss the possibility of defining set generic durations for analyzing a certain type of music clip while training. And in doing so we cut down the need of processing the clip duration to just 10\% of the global average.   

\end{abstract}
\begin{IEEEkeywords}
feature extraction, librosa, spectral features, preprocessing
\end{IEEEkeywords}
\section{Introduction}
With the music industry now booming more than ever, demand for faster and more efficient content-based music retrieval software[14] is at an all-time high. Not only music but the requirement of most efficient and accurate audio processing[13], in general, is becoming an inevitable task, be it recording and sending audio for calls, voice search[24], voice recognition[25], or music genre detection[15]. Processing audio clips, especially music, is a resource-intensive and time-consuming task. Although many higher-level abstraction APIs[1] have made a breakthrough in many music information processing tasks and using it for making better music generators[2]; when it comes to music, it is necessary to process the whole clip for accurate feature extraction. But doing so, generates a large overhead and/or becomes computationally expensive to process every single audio clip to its full length.

Here we propose a simple way of overcoming this limitation, by finding the best trims of clips to process, to extract features from it to know about the music clip. A possibility of structuring music clips to extract features from mere 5-10 seconds of clip processing, as accurately as the whole clip information extraction. To support our hypothesis, we experimented on our custom dataset, which we have documented in this paper.

The process of music information retrieval has always been a personal choice. Many music information extraction tasks can be performed using the prebuilt libraries in Python, and we used Librosa[3] for our experiments. When it comes to general genres of music, the information extraction can be done by simply taking 3-4 key attributes for understanding the pattern of the audio file. While working with more specific and relative genres, we do feature extraction by making a spectrogram of that particular audio and then extracting necessary features from it[4]. While this pattern works all the time, the computational toll of the extraction remains relatively the same. There are many datasets related to music genres, like, the SDM dataset and GTZAN dataset[5]. The main difference between the two is what we call - The ‘structure of music clip’. What we observe here, is an interesting pattern. While working with clips from SDM dataset and GTZAN, the one extracted from the SDM dataset, had to be processed as a whole to get proper information, but while processing any clip from the GTZAN dataset, we may only use just 5-10 seconds of the clip for information retrieval, and still get an accurate analysis. This led to the idea of audio analysis with smaller clips. What we wanted to see was the possibility, to format the music file, or trim the audio file in such a way that even after having processed a relatively smaller length of audio, we get the same high-quality feature extractions.

\section{Literature Review}
Holk, Eric, et al. gives an approach that seamlessly integrates with many of Rust's features, making it easy to build a library of ergonomic abstractions for data-parallel computing [1]. Conklin, Darrell, discusses the use of statistical models for the problem of musical style imitation [2]. McFee, Brian, et al. provides a brief overview of the library's functionality and provides explanations of the design goals, software development practices, and notational conventions [3]. Downie, J. Stephen, introduces the concept of music information retrieval [4]. Tzanetakis, George, and P. Cook published the GTZAN genre collection dataset [5]. Ellis, Dan, introduced the concept of chroma feature analysis and synthesis [6]. Logan, Beth, explains the feature MFCC for Music Modeling [7]. Chai, Tianfeng, and Roland R. Draxler introduced the concept of RMSE and arguments against literature [8]. Harako, Koudai, et al., explains the importance of the Roll-off factor [9].  McKinney, Martin, and Jeroen Breebaart explain the need for spectral bandwidth and centroids in feature extractions [10]. Gouyon, Fabien, François Pachet, and Olivier Delerue, justifies the use of a zero-crossing rate in the analysis [11]. Gulli, Antonio, and Sujit Pal introduce the use of Keras for Deep Learning [12].  Foote, Jonathan’s work gave us an overview of information retrieval, which then used in our experiment [13]. Tseng, Yuen-Hsien’s paper on content-based retrieval of music collections helped us better understand the music analysis process [14].Tzanetakis, George, and Perry Cook’s work on musical genre classification gave us insight on how to study and use various spectral features for our use [15]. For further advance and detailed study of spectral features and their analysis, we studied works of Rauber and M. Fruhwirth on Automatically analyzing ¨ and organizing music archives [16],  Liu and Q. Huang on Content-based indexing and retrieval-by-example in audio [17],  B. Dannenberg, B. Thom, and D. Watson on A machine learning approach to musical style recognition [18], Aucouturier, J. J., \& Pachet, F. on use of music similarity measures [19]. We studied Sharma, Garima, Kartikeyan Umapathy, and Sridhar Krishnan’s work to know recent and most efficient trends in feature extraction methods [20].  Bajaj, Anu, Tamanna Sharma, and Om Prakash Sangwan and their work on Information Retrieval in Conjunction With Deep Learning was the first paper we started our research from [21]. Saikkonen, Lauri's work on structured music and the behaviour of its spectral features gave us the idea of using it in our experiments [22]. Revilla, Melanie, et al. tested the use of voice input in smartphones, which we studied to find difference between audio and music analysis [23]. Muda, Lindasalwa, Mumtaj Begam, and Irraivan Elamvazuthi’s work on voice recognition algorithms helped us better test and develop our own music clip information retrieval algorithms [24].

\section{Methodology}

\begin{figure}[h]
\centerline{\includegraphics[width=0.7\columnwidth]{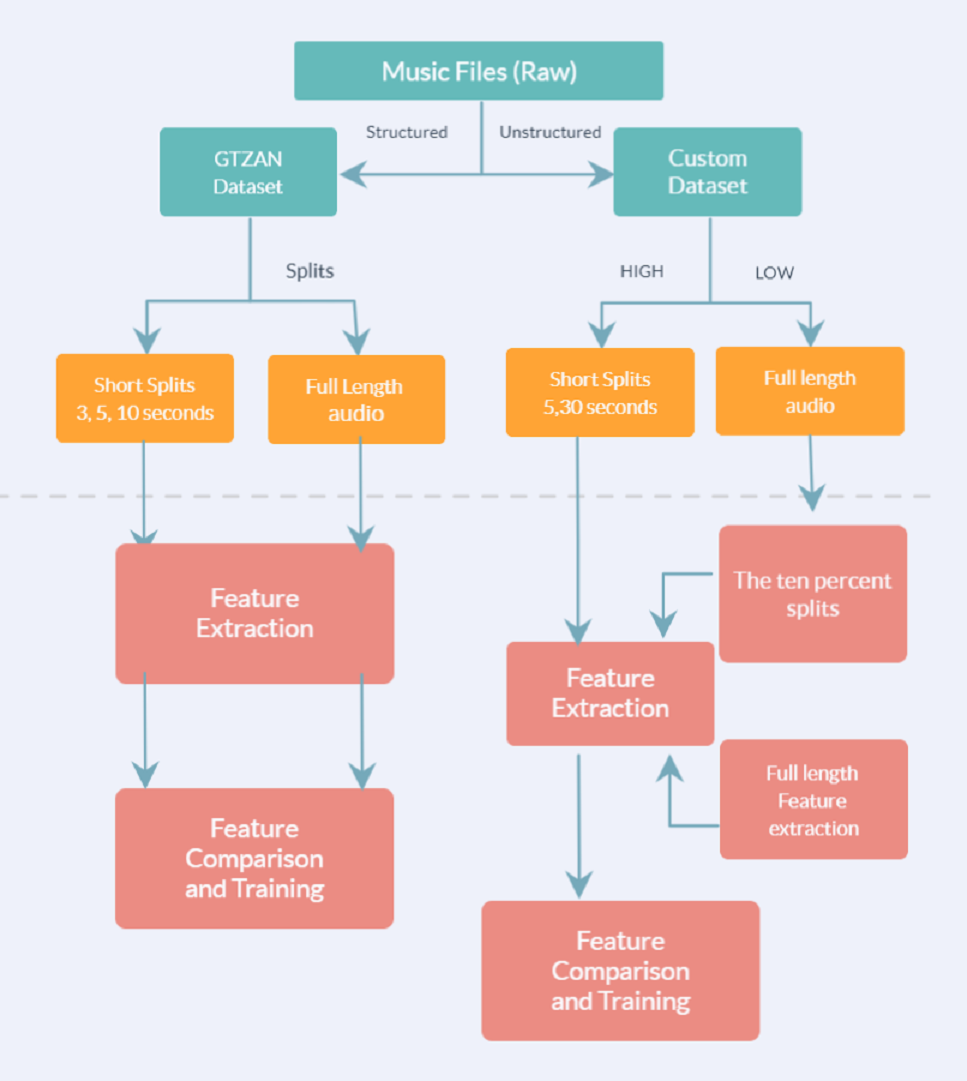}}
\caption{An overview of methodology followed.}
\label{flow}
\end{figure}

\subsection{Librosa}
For music and audio analysis, Python provides Librosa package, which we have used in this experiment for feature extraction purposes. It is used by a vast majority of the developers for the same purpose as it provides the necessary building blocks to create music information retrieval systems.

While using librosa for music information retrieval, we get many variables, also called features, as output of the process. These features are then analysed by user as per the needs. The main features we considered from the report are:

\subsubsection{Chroma\_stft}
Chroma Feature analysis [6] and synthesis is a powerful and interesting representation of music audio, in which the entire spectrum is divided into 12 bins which represent 12 distinct chromas (semitones) of the musical octave. Since in music, notes that are one octave apart are considered to be similar, knowing the chroma distribution (spectrograph), without actually knowing the absolute frequency, can give useful musical information about the audio that is not apparent even in the original spectra.

\subsubsection{MFCC}
The mel frequency cepstral coefficients (MFCCs) [7] of a signal are a small set of features (usually about 10-20) which concisely describe the overall shape of a spectral envelope. MIR is often used to describe timbre. This is mainly used for music structure analysis.

\subsubsection{RMSE}
This particular feature is used to compute root-mean-square (RMS) energy [8] for each frame, either from the audio samples y or from a spectrogram S. Here we are using the y (the audio file) for generating computations.
\subsubsection{Roll-off}
The roll-off factor [9], is a measure of the excess bandwidth of the filter, i.e. the bandwidth occupied beyond the Nyquist bandwidth. Here the roll-off is used in order to check how much excess bandwidth is passing through.

\subsubsection{Spectral bandwidth}
Spectral bandwidth [10], signifies the state of the art features with which we can easily understand the state of the audio at that particular time, after which anything relative to the audio information can be extracted easily.

\subsubsection{Spectral Centroid}
Spectral centroid [10] is more relative to spectral bandwidth and is often neglected, as the correlation between the two makes it too obvious to consider as an important feature.
\subsubsection{Zero Crossing Rate}
This feature is used to calculate the zero-crossing rate [11] of the audio clip. The reason behind the importance of this particular feature is to find out at what speed, or with how much variations the particular music clip is following through. This could be extremely helpful in identifying the genre of that particular clip.
\section{Dataset Information}
\subsection{GTZAN Dataset}\label{AA}
The dataset consists of 1000 audio tracks each 30 seconds long. It contains 10 genres, each represented by 100 tracks. The tracks are all 22050Hz Mono 16-bit audio files in .wav format. This particular dataset is the one that we call a structured music clip dataset. As the music tone and audio signal have the same kind of pattern throughout the clip, even a small piece of the clip can give about the same kind of information, necessary to discover some basic information about the music, like the genre. 
\subsection{Custom Dataset}
The SDM Dataset is a custom dataset that we created that contained over 800 unaltered songs of different genres. It contains 8 different genres with 100 songs each. This particular dataset is what we call, the unstructured one. This means that the song clip may vary from time to time, and the particular duration trimming may give the wrong set of features, which may not even represent that music clip accurately. Hence we need to process this clip for the whole duration. 

\section{Implementation Strategy}
The major part of the implementation strategy has been discussed earlier, hence, to sum up, we worked with two datasets, structured and unstructured. For feature extraction, we used a prebuilt Python audio processing library, Librosa. We followed the feature extraction cycle from making spectrograms and feature extraction, to training the model on those features to have a simple classification to test accuracy. 

\section{The Actual method}
First, we transformed our full-length music clips to spectrograms and saved them as image files, to be used for later analysis. Then, we started extracting features with Librosa, and appended them into the features file (.csv). Finally that CSV file was used to train a model using the high-level API named Keras[12]. 
In the feature extractions, each feature is extracted on instances from the audio clip. We do use a duration parameter to control the length of the clip getting scanned. The length control is important for the experiments done later on, on the unstructured datasets and the 10 percent trimmed clips. 

\section{Training}
The specific training conditions provided were:
\begin{itemize}

\item Sequential Model
\item Optimizer: Adam
\item Loss: Sparse Categorical Cross-Entropy
\item Activation: ReLU and Softmax
\item Evaluation Matrix: Accuracy
\item Batch Size: 128
\item Training samples: 1600
\item Evaluation Samples: 200
\end{itemize}
We trained the model with features extracted from 7 different audio clip lengths using  (3, 5, 10, and 30) seconds of clip lengths from the GTZAN dataset and (5, 20, and 180) seconds clip lengths from the SDM Dataset.

\section{Experiments and Discussions}

Here Figure 2 shows the chroma\_stft values extracted from the audio clips taken from the GTZAN dataset. The duration for which the audio clips were processed are as mentioned - 3, 5, 10, and 30 seconds. The same link can be found with the zero-crossing rate feature (Figure 3). These graphs show that no matter how short or long the clip is processed for, the features stay consistent. Which further supports the claim, that if the audio file is structured or if some parts are confined and seen together, the music feature information can still be extracted from even a smaller part of the original audio file. And the features match consistently with the original length analysis.

\begin{figure}[tb]
\centerline{\includegraphics[width=0.7\columnwidth]{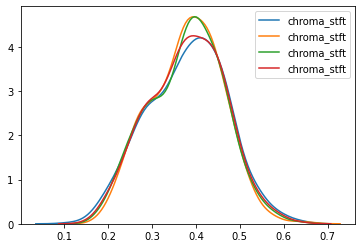}}
\caption{The clips from the structured data processed for 3, 5, 10 and 30 seconds.}
\label{fig2}
\end{figure}

\begin{figure}[h]
\centerline{\includegraphics[width=0.7\columnwidth]{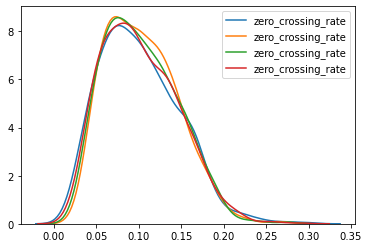}}
\caption{The clips from the structured data processed for 3, 5, 10 and 30 seconds.}
\label{fig3}
\end{figure}

Then comes the Figure 4 and Figure 5, where the audios processed, are taken from the unstructured dataset, i.e. the SDM dataset, the duration for which the audio clips were processed are as mentioned (5, 20, and 180 seconds). 
\begin{figure}[h]
\centerline{\includegraphics[width=0.7\columnwidth]{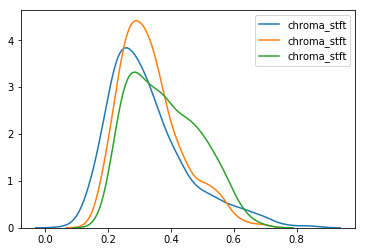}}
\caption{The clips from the unstructured data processed for 5, 20 and 180 seconds.}
\label{fig4}
\end{figure}

\begin{figure}[t]
\centerline{\includegraphics[width=0.7\columnwidth]{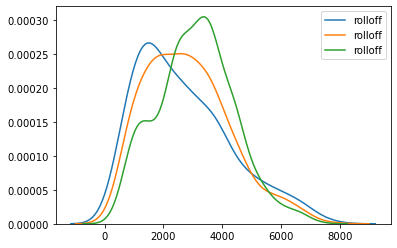}}
\caption{The clips from the unstructured data processed for 5, 20 and 180 seconds.}
\label{fig5}
\end{figure}

We observe that the features do not stay the same when we change the duration-of-processing of the clip. This is the pattern we recognized. And this holds true for all the experiments we conducted. The model was trained on the features extracted from all the length of processing and the trend seen was as follows. The model trained on features extracted from the structured dataset, on all the lengths, showed pretty similar accuracy. The results of the classification can be seen in Table 1. However, the model when trained on different accuracies of the unstructured dataset, it did change with the length of processing. The model wasn’t made efficient, because we just wanted to see the effect of duration change and if the features change or not.

Notable observations can be made from Figure 6 and Figure 7 where the graphs are shown for the features extracted from the last snippet of the global average range(i.e. 180s).

\begin{figure}[h]
\centerline{\includegraphics[width=0.7\columnwidth]{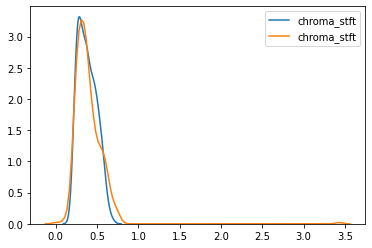}}
\caption{Comparison between 180s and last 10\% of custom clip length}
\label{fig6}
\end{figure}

\begin{figure}[b]
\centerline{\includegraphics[width=0.7\columnwidth]{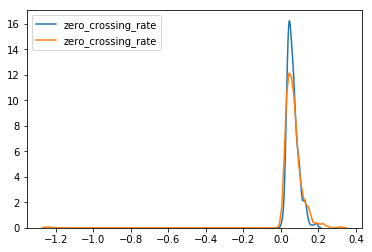}}
\caption{Comparison between 180s and last 10\% of custom clip length}
\label{fig7}
\end{figure}

These are shown with the features extracted from the whole length clip processing. They show a very similar trend, and this trend holds for all other features too. What we observe from this is that the features can be extracted from a small length even with unstructured data. What is even more important is that a simple process of knowing the audio clip is all that is needed. Here we found that only the last 10 percent of the 180 second average that we kept, is important for accurately extracting the features. Not only the graphs but the model accuracies also support the claim.

\begin{table}[h]
\centering
\caption{ Accuracy of the model classification. For further support to, the similarity of feature extraction while working with ‘structured’ dataset.}
\label{tab:table}
\begin{tabular}{@{}|l|l|l|@{}}
\toprule
\textbf{ID of Experiments} & \textbf{Duration} & \textbf{Test Accuracy} \\ \midrule
E1                         & 3s                & 0.680                   \\ \midrule
E2                         & 5s               & 0.695                  \\ \midrule
E3                         & 10s               & 0.690                   \\ \midrule
E4                         & 30s               & 0.704                  \\ \bottomrule
\end{tabular}
\end{table}

As we see in the table, we achieve the same results even in the classification model, which further supports the idea that features extracted by processing any length of the clip from the structured dataset, shows a similar trend and are consistent. But if we use the unstructured dataset, then it performs worse on small clip feature extractions. As we can see in Table 2, the results show the drastic change in the accuracy of classifications.

\begin{table}[h]
\centering
\caption{Accuracy of the model classification. For further support to, dissimilarity of feature extraction while working with the ‘unstructured’ dataset.}
\label{table 2}
\begin{tabular}{@{}|c|c|c|@{}}
\toprule
\textbf{Number of exp} & \textbf{Duration}                                                                               & \textbf{Test Accuracy} \\ \midrule
E1                     & 5s                                                                                              & 0.694                 \\ \midrule
E2                     & 20s                                                                                             & 0.7686                 \\ \midrule
E3                     & 180s                                                                                            & 0.7835                  \\ \midrule
E4                     & \begin{tabular}[c]{@{}c@{}}last 10\% \\ (162 -180s slot)\end{tabular} & 0.7849                 \\ \bottomrule
\end{tabular}
\end{table}

Here, the last result (E4), shows the accuracy of the model trained on the features extracted from the last 10 percent split of the 180 seconds average. We see that the testing accuracy holds true and similar to the model trained with features extracted from the whole length of the clip.

This is what changes the whole process of feature extraction, especially with the music audio clips. When we talk about the clips now, we can easily extract the spectral features by simply taking the 162-180 second slot time of the music clip and the features will match pretty well with the full length audio too. And not just that, it would also not hinder the model accuracies if made. 
\section{Future Scope and Conclusion}

As we know that the structuring and formatting of the audio clip could lead to more efficient feature extraction, we can now develop a model to test the best time to trim the clip, and for how much length, to efficiently extract the features, by just 15 to 20 seconds of audio length to be processed. This can be done by making a bigger dataset with defined labels, and then dividing the clips into many different length clips. After that, the features can be extracted for that length and compared with the total length processing. This would lead to a conclusive result that at which length, which genre is at its optimum and can give the near exact feature extraction. This would then lead to greater improvements to the efficiency of the feature extraction process. This can take the computational toll down by at least 10 times the current value. When it comes to the use of the technology, this method can be used for faster analysis in the music applications, which deals with the features like understanding the likings of a particular user. For that most applications use data analysis on two features, one is the lyrics and second the tune, i.e. the spectral features. By considering the flow of music into such applications every day, if someone needs to give somewhat of an analysis every time a new song is entered, it can be done dynamically by using our technique. The workaround is going to lead to near-perfect spectral feature extraction, at least 10 times faster.


\begin{thebibliography}{00}
\bibitem{b1} Holk, Eric, et al. "GPU programming in rust: Implementing high-level abstractions in a systems-level language." 2013 IEEE International Symposium on Parallel \& Distributed Processing, Workshops and Phd Forum. IEEE, 2013.

\bibitem{b2}  Conklin, Darrell. "Music generation from statistical models." Proceedings of the AISB 2003 Symposium on Artificial Intelligence and Creativity in the Arts and Sciences. 2003.


\bibitem{b3}McFee, Brian, et al. "librosa: Audio and music signal analysis in python." Proceedings of the 14th python in science conference. Vol. 8. 2015.

\bibitem{b4}Downie, J. Stephen. "Music information retrieval." Annual review of information science and technology 37.1 (2003): 295-340.


\bibitem{b5} Tzanetakis, George, and P. Cook. "GTZAN genre collection." Music Analysis, Retrieval and Synthesis for Audio Signals (2002).
\bibitem{b6} ] Ellis, Dan. "Chroma feature analysis and synthesis." Resources of Laboratory for the Recognition and Organization of Speech and Audio-LabROSA (2007).

\bibitem{b7} Logan, Beth. "Mel Frequency Cepstral Coefficients for Music Modeling." ISMIR. Vol. 270. 2000.


\bibitem{b8}  Chai, Tianfeng, and Roland R. Draxler. "Root mean square error (RMSE) or mean absolute error (MAE)?–Arguments against avoiding RMSE in the literature." Geoscientific model development 7.3 (2014): 1247-1250.

\bibitem{b9} Harako, Koudai, et al. "Roll-off factor dependence of Nyquist pulse transmission." Optics express 24.19 (2016): 21986-21994.


\bibitem{b10}  McKinney, Martin, and Jeroen Breebaart. "Features for audio and music classification." (2003).

\bibitem{b11}Gouyon, Fabien, François Pachet, and Olivier Delerue. "On the use of zero-crossing rate for an application of classification of percussive sounds." Proceedings of the COST G-6 conference on Digital Audio Effects (DAFX-00), Verona, Italy. 2000.


\bibitem{b12}Gulli, Antonio, and Sujit Pal. Deep Learning with Keras. Packt Publishing Ltd, 2017.

\bibitem{b13} Foote, Jonathan. "An overview of audio information retrieval." Multimedia systems 7.1 (1999): 2-10. 


\bibitem{b14}Tseng, Yuen-Hsien. "Content-based retrieval for music collections." Proceedings of the 22nd annual international ACM SIGIR conference on Research and development in information retrieval. 1999.


\bibitem{b15}Tzanetakis, George, and Perry Cook. "Musical genre classification of audio signals." IEEE Transactions on speech and audio processing 10.5 (2002): 293-302.


\bibitem{b16}Rauber, Andreas, and Markus Frühwirth. "Automatically analyzing and organizing music archives." International Conference on Theory and Practice of Digital Libraries. Springer, Berlin, Heidelberg, 2001.


\bibitem{b17}Liu, Zhu, and Qian Huang. "Content-based indexing and retrieval-by-example in audio." 2000 IEEE International Conference on Multimedia and Expo. ICME2000. Proceedings. Latest Advances in the Fast Changing World of Multimedia (Cat. No. 00TH8532). Vol. 2. IEEE, 2000.


\bibitem{b18}Dannenberg, Roger B., Belinda Thom, and David Watson. "A machine learning approach to musical style recognition." (1997).


\bibitem{b19}Aucouturier, Jean-Julien, and Francois Pachet. "Music similarity measures: What's the use?." ISMIR. 2002.



\bibitem{b20}Sharma, Garima, Kartikeyan Umapathy, and Sridhar Krishnan. "Trends in audio signal feature extraction methods." Applied Acoustics 158 (2020): 107020.

\bibitem{b21}]Bajaj, Anu, Tamanna Sharma, and Om Prakash Sangwan. "Information Retrieval in Conjunction With Deep Learning." Handbook of Research on Emerging Trends and Applications of Machine Learning. IGI Global, 2020. 300-311.


\bibitem{b22}Saikkonen, Lauri. "Structural analysis of recorded music." (2020).

\bibitem{b23}Revilla, Melanie, et al. "Testing the use of voice input in a smartphone web survey." Social Science Computer Review 38.2 (2020): 207-224.


\bibitem{b24}Muda, Lindasalwa, Mumtaj Begam, and Irraivan Elamvazuthi. "Voice recognition algorithms using mel frequency cepstral coefficient (MFCC) and dynamic time warping (DTW) techniques." arXiv preprint arXiv:1003.4083 (2010).


\end{thebibliography}
\end{document}